\documentclass[ onecolumn, 12pt]{IEEEtran}
\usepackage {makeidx, ifthen}

\usepackage{graphicx}   
\usepackage[center]{caption2}

\usepackage{amsmath}
\usepackage{amssymb}
\usepackage{latexsym}
\usepackage {makeidx, ifthen}

\usepackage{array}
\newtheorem{theorem}{Theorem}

\newtheorem{definition}{Definition}
\newtheorem{lemma}{Lemma}

\begin{document}

\title{An Improved Lower Bound to the Number of Neighbors Required for the Asymptotic Connectivity of Ad Hoc Networks}


\author{\authorblockN{Sanquan Song\thanks{Sanquan Song is with the
Department Electrical Engineering and Computer Science,
Massachusetts Institute of Technology, Cambridge, MA 02139. (email:
sanquan@mit.edu).}, Dennis L. Goeckel\thanks{Dennis L. Goeckel is
with the Department of Electrical and Computer Engineering,
 University of Massachusetts Amherst, Amherst, MA 01002 (email: goeckel@ecs.umass.edu).},
Don Towsley\thanks{Don Towsley is with the Department of Computer
Science,  University of Massachusetts Amherst, Amherst, MA 01002
(email: towsley@cs.umass.edu).}}
\thanks{Communicated by Sanquan Song.}
\thanks{This paper is based upon work supported in part by the
     Army Research Office under Contract DAAD10-01-1-0477
     and the National Science Foundation under grants
     ECS-0300130 and CCF-0430892, and employed equipment
     obtained under National Science Foundation grant
     EIA-0080119.}
}


\maketitle

\begin{abstract}
Xue and Kumar \cite{1Xue and Kumar: Number} have established that
the number of neighbors required for connectivity of wireless
networks must grow as $\Theta(\log N)$, and \cite{1Xue and Kumar:
Number} also established that the actual number required lies
between 0.074 $\log N$ and 5.1774 $\log N$.  In this short paper, by
recognizing that connectivity results for networks where the nodes
are distributed according to a Poisson point process can often be
applied to the problem of [3], we are able to improve the lower
bound. In particular, we show that a network with nodes distributed
in a unit square according to a 2D Poisson point process of
parameter $N$ will be asymptotically disconnected with probability
one if the number of neighbors is less than 0.129 $\log N$.
Moreover, $ 0.129\log\left( N+\frac{\pi}{4} -\sqrt{\frac{\pi
N}{2}+\frac{\pi ^2}{16}}\right)$ is not enough for an asymptotically
connected network with $N$ nodes uniformly in a unit square, hence
improving the lower bound from \cite{1Xue and Kumar: Number}.
\end{abstract}
\begin{keywords}
Wireless networks, ad hoc networks, connectivity, power control.
\end{keywords}


\section{Introduction}

Due to their widespread applicability, wireless ad hoc networks have
attracted significant research interest in recent years. In an ad
hoc network, each node is connected with several nearby nodes
directly and thus to others by relay via these neighbors; thus, the
neighbors of each node eventually determine the network
connectivity. Based on this observation, researchers have defined
the $k$-neighbor network model by assuming that each node adjusts
its power to maintain a link with its $k$ closest neighbors
\cite{1Xue and Kumar: Number}, and then studied network connectivity
performance as a function of $k$.

As $k$ increases, network connectivity improves. For a network with
$N$ nodes, if $k=N-1$, any pair of nodes can communicate directly,
which is the best achievable connectivity. However, node power must
increase to achieve such connectivity, which leads to more signal
interference and lower network capacity \cite{1Gupta and Kumar:
Capacity}. Thus, given the requirement that the network be
connected, the minimal $k$ that provides such is desired.

Researchers used to believe that there exists a ``magic number''
such as $k=6$, $k=8$ or $k=3$, that leads to good network
connectivity \cite{magic1}\cite{magic2}\cite{magic3}\cite{magic4}.
While such a number might be sufficient for connectivity of
small-scale networks, Xue and Kumar \cite{1Xue and Kumar: Number}
find that a large-scale network is disconnected with probability one
when a fixed $k$ is employed. They study the analogous problem in
the dense network case \cite{1Xue and Kumar: Number} and show that,
if there are $N$ nodes uniformly located in a unit square, each node
should be connected with $\Theta (\log N)$ nearest neighbors so that
the network is connected with probability one asymptotically as $N$
goes to infinity. The exact value of $k$ that guarantees the
connectivity should be more than $0.074\log N$ and less than
$5.1774\log N$.

In addition to the $k$-neighbor model, there exists the $r$-radius
 model for a wireless ad hoc network, where all nodes employ
the same radio power, and thus each node can establish a direct link
with any other node within some fixed distance $r$. Gupta and Kumar
\cite{1Gupta and Kumar: Connectivity} find that in a network with
$N$ nodes uniformly distributed in a unit area disk, the network is
connected with probability one as $N\to\infty$ if and only if $ \pi
r^2=\frac{ \log N+c(N)}{N}$ and $\lim\limits_{N\to\infty}
c(N)=\infty$. Naturally, in this case, the expected number of
neighbors of one node is $ N\pi r^2= {\log N+c(N)}$. Comparing this
result with that for the $k$-neighbor model, it is reasonable to
conjecture that the true value of $k$ should take the form of $\log
N+c(N)$. However, this has not been established.

The argument in \cite{1Xue and Kumar: Number} leading to the lower
bound of $0.074\log N$ is very complicated. The most important
reason for this complexity is the dependence of the nodes placed in
two non-overlapping areas for a network consisting of $N$ nodes
uniformly distributed in a unit square (denoted by $G(N)$ here).
However, such a dependence does not exist in a network
$G^{Poisson}(N)$, where the total number of nodes is a Poisson
random variable with parameter $N$. Inspired by Lemma 4 in
\cite{ours}, it is possible to study the connectivity performance of
the $k$-neighbor network $G^{Poisson}(N)$ and then establish a link
from $G^{Poisson}(N)$ to $G(N)$.  In this paper, we use this
approach to improve the lower bound on the number of neighbors
required for the asymptotic connectivity of ad hoc networks.


\section{$0.129\log N$ neighbors are necessary for connectivity} 
We focus on $G^{Poisson}(N)$, where nodes are distributed according
to a two-dimensional Poisson point process. First, a disconnection
pattern for the network is defined. Then the probability of there
existing at least one such pattern is studied and a lower bound to
this probability is obtained, which is a function of $k$. Thus, for
any $k$ that makes this lower bound go to one, a lower bound of $k$
below which the network will be disconnected asymptotically with
probability one is obtained. Finally, a link is made from
$G^{Poisson}(N)$ to $G(N)$ that yields the desired result.

\subsection{A Scenario for Disconnection}
\begin{figure}[htbp]
\centering
\includegraphics[ width=3.0in ] {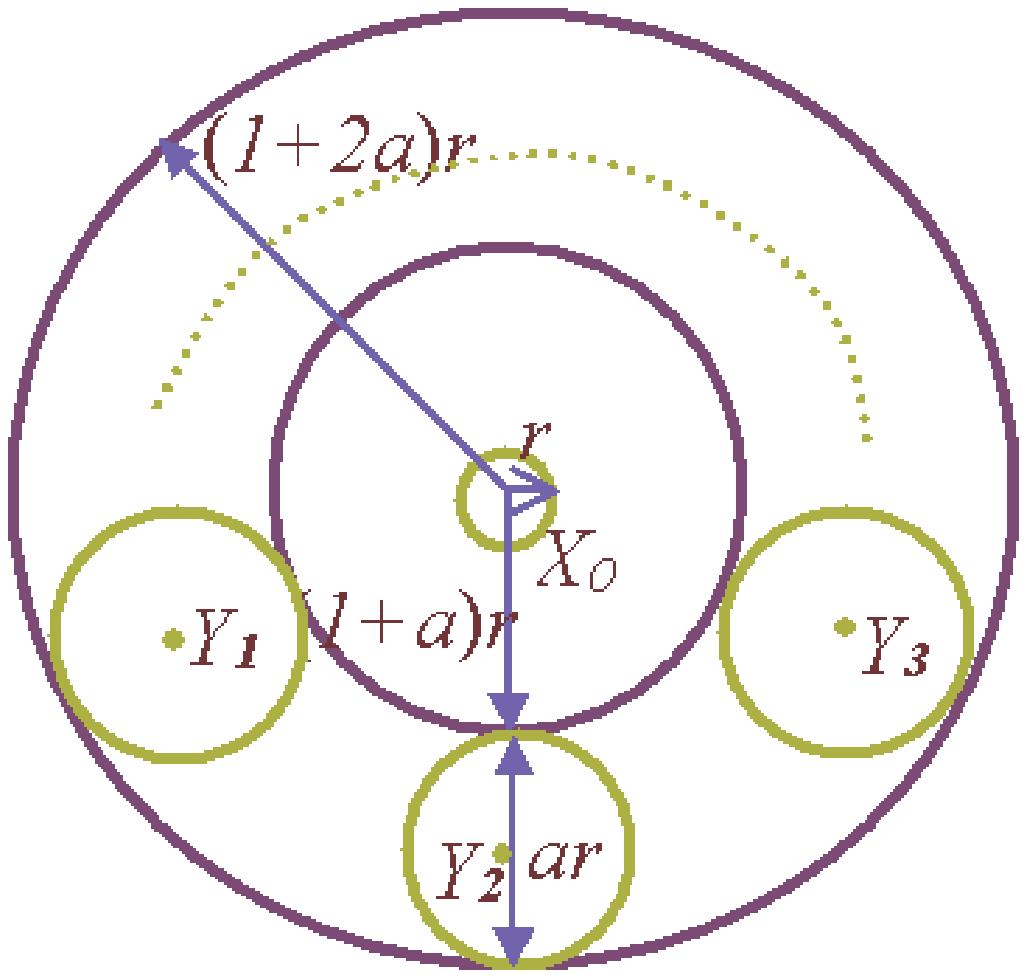}
\caption{Trap of $d(r,a, L)$.} \label{fig: demo of trap}
\end{figure}

\begin{definition}
$B_r(X)$: a disk centered at $X$ with radius $r$.
\end{definition}
\begin{definition}
Trap of type $d(r,a,L)$: a structure with three disks centered at
the same point $X_0$, namely $B_r(X_0)$, $B_{(1+a)r}(X_0)$ and
$B_{(1+2a)r}(X_0)$ (see Figure \ref{fig: demo of trap}).
Furthermore, $L$ non-overlapping disks of radius $ar/2$ (denoted by
$B_{ar/2}(Y_i)$, $i\in[1,L]$) are evenly spaced in the annulus of
inner radius $(1+a)r$ and outer radius $(1+2a)r$. We call this
structure a trap of type $d(r,a, L)$.
\end{definition}

Such a structure will cause a disconnection when the nodes are
distributed according to some rules and the parameter $L$ is large
enough, thereby motivating the name `{\emph {trap}}'.

\begin{figure}[htbp]
\centering
\includegraphics[ width=3.0in ] {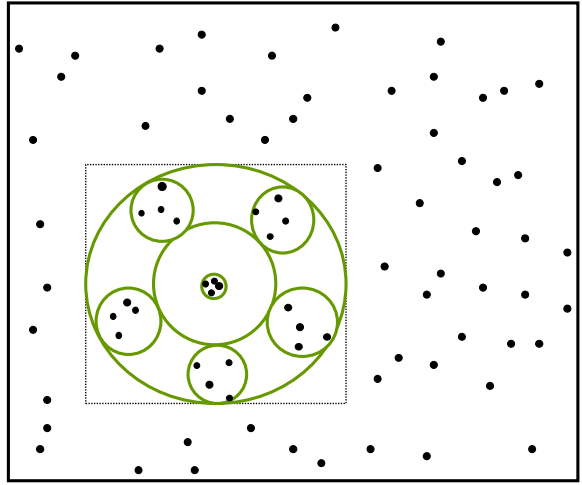}
\caption{A $k$-filling event for a trap of type $d(r,a,L)$, where
$k=4$ and $L=5$.} \label{fig: k-filling}
\end{figure}

\begin{definition}
$k$-filling event:  a trap of type $d(r,a, L)$, where there exist
$k$ nodes in the disk $B_r(X_0)$ and in each of the disks
$B_{ar/2}(Y_i)$, $i\in \{1,\dots\, L\}$,
 and no additional nodes elsewhere in the disk
$B_{(1+2a)r}(X_0)$ (See Figure \ref{fig: k-filling}).
\end{definition}

Consider the number of non-overlapping disks of diameter $ar$ that
can be placed in the annulus of inner radius $(1+a)r$ and outer
radius $(1+2a)r$.
\begin{lemma}
Let $L_{max}(a)$ denote the maximum value that $L$ can take for a
trap of type $d(r, a, L)$. Then $L_{max}(a)$ is a function of $a$
given by:
\begin{align}
L_{max}(a)=\left \lfloor \frac{\pi}{\arcsin(a/(2+3a))} \right
\rfloor
\end{align}
\end{lemma}
\begin{proof}
See details in Figure \ref{fig: L max}.

\begin{figure}[htbp]
\centering
\includegraphics[ width=3.5in ] {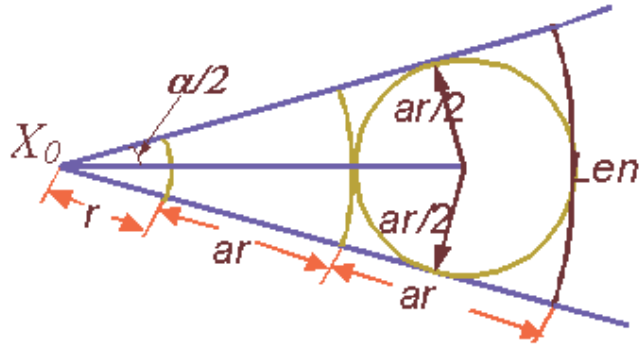}
\caption{Estimation of $L_{max}(a)$.} \label{fig: L max}
\end{figure}

\begin{align}
\alpha &= 2\arcsin\left(\frac{ar/2}{(1+3a/2)r}\right)=
2\arcsin\left(\frac{a}{2+3a}\right)\label{alpha}\\
 &\Rightarrow L \leq \left \lfloor \frac{\pi}{\arcsin(a/(2+3a))} \right\rfloor
\end{align}
\end{proof}

\begin{lemma}
For a given $a$, $\exists L_{min}(a)\leq L_{max}(a)$ such that
$\forall L\in\left[L_{min}(a), L_{max}(a)\right]$, the $k$-filling
event that occurs in a trap of type $d(r,a, L)$ implies that the
nodes in the center disk $B_r(X_0)$ of this trap are disconnected
with the nodes outside $B_r(X_0)$ and, hence, the network is
disconnected. (See Figure \ref{fig: L max2})
\end{lemma}
\begin{proof}
\begin{figure}[htbp]
\centering
\includegraphics[width=4.0in] {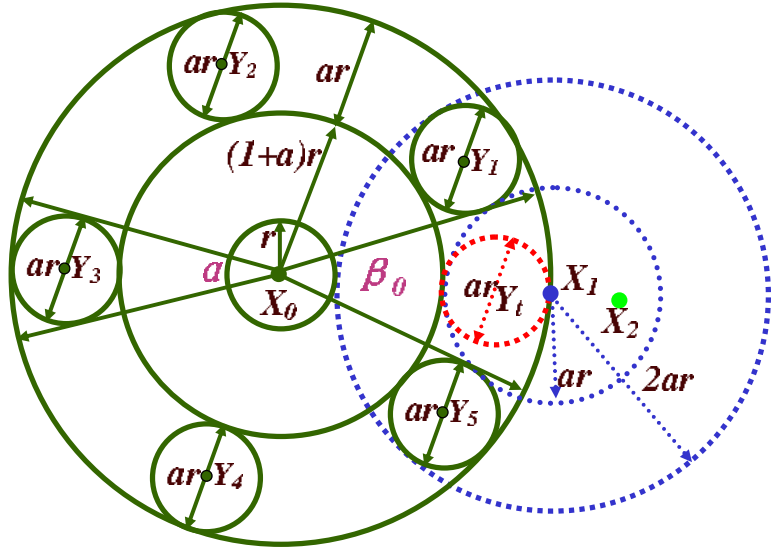}
\caption{Estimation of $L_{min}(a)$ ($L_{0}=5$ in this case).}
\label{fig: L max2}
\end{figure}

See Figure \ref{fig: L max2}. Consider a node $X_2$ which lies
outside of the disk $B_{(1+2a)r}(X_0)$. $X_2$ will choose $k$
nearest nodes as its neighbors. If it selects one of the nodes in
$B_r(X_0)$, a link from $B_r(X_0)$ to the outside of
$B_{(1+2a)r}(X_0)$ exists, and the $k$-filling event for this trap
of type $d(r,a,L)$ does not imply a disconnection scenario.
Therefore, in order to guarantee that the $k$-filling event for this
specified structure leads to a disconnection, we need to increase
$L$, the number of the sub-disks in the annulus, so that a disk
centered outside of $B_{(1+2a)r}(X_0)$ and tangent with $B_r(X_0)$
must contain at least one of the sub-disks, say $B_{ar/2}(Y_i)$
entirely. This guarantees that each node not in $B_{(1+2a)r}(X_0)$
contains a sufficient number of nodes ($\geq k$) closer to it than
any node in $B_r(X_0)$. Furthermore, we just need to find the value
of $L$ that is large enough so that any disk $B_{2ar}(X_1)$, which
is centered on the boundary of $B_{(1+2a)r}(X_0)$, must contain at
least one $B_{ar/2}(Y_i)$.

From (\ref{alpha}), $\alpha$ is fixed given $a$. Let $\beta$ denote
the angle between two neighbors $B_{ar/2}(Y_i)$ and
$B_{ar/2}(Y_{i+1})$.  Let $\beta_0$ denote the corresponding $\beta$
satisfying the condition that $B_{2ar}(X_1)$ is tangent with
$B_{ar/2}(Y_1)$ and $B_{ar/2}(Y_{5})$. Furthermore, for
$\beta=\beta_0$, $B_{ar}(X_1)$ is tangent with $B_{ar/2}(Y_1)$,
$B_{ar/2}(Y_5)$ and $B_{(1+a)r}(X_0)$ (see Figure \ref{fig: L
max2}). Therefore, we can put a disk of radius $ar/2$ in
$B_{ar}(X_1)$, denoted as $B_{ar/2}(Y_t)$ in Figure \ref{fig: L
max2}, that is tangent with $B_{(1+a)r}(X_0)$ and $B_{(1+2a)r}(X_0)$
but does not contact $B_{ar/2}(Y_1)$ and $B_{ar/2}(Y_5)$. From above
analysis, we know that $\beta_0>\alpha$ for any $a$. Obviously,
given the condition that $\beta\leq\beta_0$, the $k$-filling event
that occurs in the corresponding trap of type $d(r,a, L(a))$ implies
that the nodes in the center disk $B_r(X_0)$ of this trap are
disconnected with the nodes outside $B_r(X_0)$ and, hence, the
network is disconnected. Thus, for any $L(a)$ such that
\begin{align}
L(a)\geq  \left \lceil \frac{\pi}{2\arcsin( \frac{a}{2+3a}) } \right
\rceil =   \left \lceil \frac{2\pi}{2\alpha} \right  \rceil \geq
\left \lceil \frac{2\pi}{\alpha+\beta_{0}} \right \rceil
\end{align}
the corresponding $\beta$ is no greater than $\beta_0$. Therefore,
the $k$-filling event that occurs in the trap of type $d(r,a,
L(a))$, $L(a)\geq \left \lceil \frac{\pi}{2\arcsin( \frac{a}{2+3a})
} \right \rceil$, implies that the network is disconnected. Thus, it
yields:
\begin{align}
L_{min}(a)\leq \left \lceil \frac{\pi}{2\arcsin( \frac{a}{2+3a}) }
\right \rceil
\end{align}
which is less than $L_{max}(a)=\left \lfloor
\frac{\pi}{\arcsin(a/(2+3a))} \right \rfloor$.
\end{proof}
\begin{lemma}\label{L: division}
The maximum number $S$ of non-overlapping traps of type $d(r,a, L)$
that can be placed in a unit square is given by:
\begin{align}
S=\left \lfloor \frac{1}{2(1+2a)r} \right \rfloor^2
\end{align}
\end{lemma}
\begin{proof}
Clearly, we can divide the square into sub-squares of edge length
$2(1+2a)r$ and put one trap into one sub-square. So we have:
\begin{align}
S=\left \lfloor \frac{1}{2(1+2a)r} \right \rfloor^2
\end{align}
\end{proof}

\begin{lemma}\label{L: k-filling}
In $G^{Poisson}(N)$, the probability of a $k$-filling event for a
trap of type $d(r,a, L)$ is:
\begin{align}
P_{k-filling}=\frac{(N\pi r^2)^k}{k!}\left [  \frac{(N\pi
a^2r^2/4)^k}{k!}\right ]^Le^{-N\pi (1+2a)^2r^2}
\end{align}
\end{lemma}
\begin{proof}
A $k$-filling event for a trap of type $d(r,a, L)$ means that there
are $k$ nodes in each of the disks $B_r(X_0)$ and $B_{ar/2}(Y_i)$,
$i\in\{1,\dots L\}$ and no additional node elsewhere in the disk
$B_{(1+2a)r}(X_0)$. Therefore:
\begin{align}
P_{k-filling} &= \frac{(N\pi r^2)^k}{k!}e^{-N\pi r^2} e^{-N(\pi
(1+2a)^2r^2-\pi r^2-L\pi a^2r^2/4)}\left [ \frac{(N\pi
a^2r^2/4)^k}{k!}e^{-N\pi a^2r^2/4} \right ]^L\\
 &= \frac{(N\pi r^2)^k}{k!}\left [  \frac{(N\pi
a^2r^2/4)^k}{k!}\right ]^Le^{-N\pi (1+2a)^2r^2}
\end{align}
\end{proof}

\subsection{Probability for the Disconnection }

\subsubsection{Disconnection in $G^{Poisson}(N)$}

From the analysis in the last section, we know that the existence of
a $k$-filling event for a trap of type $d(r,a,L)$, $L>L_{min}(a)$,
in the network means that the network is disconnected. Furthermore,
if the probability of existing at least one such $k$-filling event
in the network is one, the network is disconnected with probability
one.

Divide the unit square into non-overlapping sub-squares of edge
length $2(1+2a)r$ and put one trap of type $d(r,a, L)$ into each
sub-square. As $r, a$ and $L$ are free parameters, which do not
influence the real network connectivity, we can choose any $r, a$
and $L$ to facilitate the proof. In the following derivations, $r
\downarrow 0$ as $N \uparrow \infty$; $a$ will be a constant as $N
\uparrow \infty$; $L$ is a function of $a$ and therefore it will
vary between $L_{min}$ and $L_{max}$. Naturally, the feasible number
of traps in the square $S\uparrow \infty$ as $N\uparrow\infty$ since
$r\downarrow 0$. Furthermore, we assume that the expected number of
nodes in $B_r(X_0)$ of a trap $N\pi r^2$ goes to infinity as $N
\uparrow \infty$. Since the number of nodes in $B_r(X_0)$ is a
Poisson random variable with a parameter increasing to infinity, the
probability for it to be some number $k$ shrinks to zero regardless
of the value of $k$. Therefore, the probability of a $k$-filling
event of a trap, denoted by $P_{k-filling}$, shrinks to zero as
$N\uparrow \infty$. If a network $G^{Poisson}(N)$ is connected,
there does not exist a $k$-filling event when we place $S$
non-overlapping traps of type $d(r,a,L)$ into it. Since the
distributions of nodes in separate traps are independent, the
probability that the network $G^{Poisson}(N)$ is connected is
bounded by:
\begin{align}
P_{Connected}^{Poisson} \leq (1-P_{k-filling})^S \rightarrow e^{-S
P_{k-filling}}\label{11}
\end{align}

From the above derivations, we can see that if $S P_{k-filling}
\rightarrow \infty$, the network will be disconnected with
probability one asymptotically. Hence consider:
\begin{align}
SP_{k-filling}= \left \lfloor \frac{1}{2(1+2a)r} \right
\rfloor^2\frac{(N\pi r^2)^k}{k!}\left [  \frac{(N\pi
a^2r^2/4)^k}{k!}\right ]^Le^{-N\pi (1+2a)^2r^2}
\end{align}
Considering the same expression with the floor function replaced
yields:
\begin{align}
f(a,r) &=  \left [ \frac{1}{2(1+2a)r} \right ]^2\frac{(N\pi
r^2)^k}{k!}\left [  \frac{(N\pi a^2r^2/4)^k}{k!}\right ]^Le^{-N\pi
(1+2a)^2r^2} \label{f fucntion}\\
 &= \left [ \frac{1}{2(1+2a)} \right ]^2\frac{(N\pi
)^k}{k!}\left [  \frac{(N\pi a^2/4)^k}{k!}\right
]^L(r^2)^{k(L+1)-1}e^{-N\pi (1+2a)^2r^2},
\end{align}
and,  as $r\downarrow 0$:
\begin{align} SP_{k-filling}=
f(a,r)-o(f(a,r))\label{12}
\end{align}
Hence, it is sufficient to study $f(a,r)$. Recalling
(\ref{11})(\ref{12}), we can maximize $f(a,r)$ by selecting
appropriate values of $a$, $L$, $c$, $k$, and $r$ to obtain a
tighter upper bound of $P_{Connected}^{Poisson}$. Therefore, letting
$\frac{\partial f(a,r)}{\partial (r^2)}=0$ yields $
r^2=\frac{k(L+1)-1}{N\pi (1+2a)^2}$. Substituting $
r^2=\frac{k(L+1)}{N\pi (1+2a)^2}$ into $f(a,r)$ and assuming
$k\uparrow \infty$ as $N\uparrow\infty$ yields:
\begin{align}
f\left(a,\left[\frac{k(L+1)}{N\pi (1+2a)^2}\right]^{1/2}\right) &=
\frac{1}{4(1+2a)^2}\frac{N^{kL+k}\pi^{kL+k}a^{2kL}}{(k!)^{L+1}4^{kL}}
\frac{k^{kL+k-1}(L+1)^{kL+k-1}e^{-kL-k}}{N^{kL+k-1}\pi^{kL+k-1}(1+2a)^{2k(L+1)-2}}\\
 &=\frac{1}{4}\frac{N\pi}{kL+k}\frac{k^{kL+k}(L+1)^{kL+k}a^{2kL}}{4^{kL}(1+2a)^{2k(L+1)}}\frac{e^{-kL-k}}{(k!)^{L+1}}\\
&\geq  \frac{1}{4}\frac{N\pi}{kL+k}\frac{k^{kL+k}(L+1)^{kL+k}a^{2kL}}{4^{kL}(1+2a)^{2k(L+1)}}\frac{e^{-kL-k}}{(\sqrt{2\pi}k^{k+1/2}e^{-k+1/12k})^{L+1}}\\
&= \frac{1}{4}\frac{N\pi}{(2\pi)^{(L+1)/2}(L+1)k^{(L+3)/2}}
\frac{(L+1)^{k(L+1)}a^{2kL}}{4^{kL}(1+2a)^{2k(L+1)}}e^{-\frac{L+1}{12k}}
\end{align}
Assuming that $k=c\log N$ and that $a,L, c$ are constants yields:
\begin{align}
f\left(a,\left[\frac{k(L+1)}{N\pi (1+2a)^2}\right]^{1/2}\right)
&\geq \frac{1}{4}\frac{N\pi e^{-\frac{L+1}{12c\log
N}}}{(2\pi)^{(L+1)/2}(L+1)(c\log
N)^{(L+3)/2}}\left(\frac{(L+1)^{L+1}a^{2L}}{4^L(1+2a)^{2(L+1)}}
\right)^{c\log N}\\
&= \Theta\left(\frac{N}{(c\log N)^{(L+3)/2}}\right)N^{c\log
\frac{(L+1)^{L+1}a^{2L}}{4^L(1+2a)^{2(L+1)}}}
\end{align}
Let $ y=c\log \frac{(L+1)^{L+1}a^{2L}}{4^L(1+2a)^{2(L+1)}}$. We
claim that if $y>-1$,  $f(a,r)$ goes to infinity, because:
\begin{align}
f &\geq \Theta\left(\frac{N*N^y}{(c\log N)^{(L+3)/2}}\right )\\
  &=  \Theta\left(\frac{N^{1+y}}{(c\log N)^{(L+3)/2}}\right )\\
   &\rightarrow \infty \text{ for  $y>-1$}
\end{align}
From $y>-1$, we can find easily:
\begin{align}
c < \left(- \log
\frac{(L+1)^{L+1}a^{2L}}{4^L(1+2a)^{2(L+1)}}\right)^{-1}\mbox{\hspace{0.5cm}where
} a\in(0,\infty), L\in [L_{min}, L_{max}] \label{for c}
\end{align}

Since the probability for a disconnected network is constructive,
for any $a, r$ and $L$ that fit the assumptions, the $c$ obtained in
(\ref{for c}) guarantees disconnectivity with probability one
asymptotically if $k=c \log N$. Thus, by exhaustive numerical
search, we can select $a$ and $L$ to maximize $c$. The result is
given by: $a=3.6, L_{max}=11, L_{min}\leq 6, L=11$, $c< 0.129$. Thus
if $k<0.129\log N$, $f(a,r)$ will go to infinity, and, hence,
$SP_{k-filling}$ will go to infinity. Since $
P_{Connected}^{Poisson} \approx e^{-SP_{k-filling}}$, $
P_{Connected}^{Poisson}$ will go to zero as $N\uparrow \infty$.

In this way, we obtain:
\begin{theorem}\label{T: Poisson}
For a $G^{Poisson}(N)$, if each node connects with $k$ nearest
neighbors, where $k< 0.129\log N$, the network will be disconnected
with probability one as $N$ goes to infinity.
\end{theorem}

In this part,  $a$ and $L$ are  two parameters used to bound $
P_{Connected}^{Poisson}$. It is natural to conjecture that any kind
of square divisions and any kind of traps can be applied to bound
the probability for network connection. In order to minimize
$f(a,r)$ in (\ref{f fucntion}) , we set $ N\pi (1+2a)^2r^2=k(L+1)$,
which is just the expected number of nodes in a $B_{(1+2a)r}(X_0)$.
We achieve the maximal value of $c$ when $L=L_{max}$ because the
expected number of nodes in a trap increases as $N\uparrow \infty$.
So a trap of type $d(a,r,L)$ is likely to have less empty area and
therefore as many $L$ sub-disks as possible. The model of a
$k$-filling event can be extended by allowing more than $k$ nodes in
the disk $B_r(X_0)$ and $B_{ar/2}(Y_i)$ (see Figure \ref{fig: L
max2}). This extension represents a more general case of
disconnection than the original $k$-filling event and might be
useful for improving the bound further.

\subsubsection{Disconnection in $G(N)$}

One of the most important differences between $G(N)$ and
$G^{Poisson}(N)$ is that, due to the fixed total number of nodes,
the probability of the $k$-filling events of non-overlapping traps
are not independent in $G(N)$, which introduces technical
difficulties if we directly apply the method above developed for
$G^{Poisson}(N)$. It is the dependence of these events that is the
reason for the complication of \cite{1Xue and Kumar: Number}. Now,
we want to find a connection between $G^{Poisson}(N)$ and $G(N)$ so
that Theorem \ref{T: Poisson} can be applied.

Let $h(N)$ be the actual number of nodes in the unit square for
$G^{Poisson}(N)$. Obviously, $h(N)$ is a Poisson distributed random
variable with parameter $N$. Recall Lemma 4 of \cite{ours}:
\begin{align}
 \lim\limits_{N\to \infty}P\left(h(N): h(N)\in\left[N-\sqrt{\pi N/2}, N+\sqrt{\pi N/2}\right]\right)=1
\end{align}
Thus as $N$ increases to infinity, although the true number of nodes
in the square is a random variable, the ratio between the
fluctuation $\sqrt{\pi N/2}$ and $N$ goes to zero with probability
one. Thus, it is natural to infer that Theorem \ref{T: Poisson} is
also correct for $G(N)$.

Here is the brief idea of the following proof: we just assume that a
network $G(N+\sqrt{\pi N/2})$ with $k<0.129 \log N$ will  be
connected with strictly positive probability as $N\uparrow \infty$.
Then we find that for any $h(N)\in [N-\sqrt{\pi N/2},N+\sqrt{\pi
N/2}]$, the network $G(h(N))$ with $k<0.129 \log N$ will be
connected with strictly positive probability, too. Therefore, we get
a conclusion that a network $G^{Poisson}(N)$ with $k<0.129 \log N$
will be connected with strictly positive probability, which
contradicts Theorem \ref{T: Poisson}. So the assumption is
incorrect, and the following theorem arises:

\begin{theorem}
\label{T: Poisson_not}
 For a network $G(N)$ in a unit square, if
each node connects with $k_0(N)$ nearest neighbors, where $ k_0(N) <
0.129\log\left( N+\frac{\pi}{4} -\sqrt{\frac{\pi N}{2}+\frac{\pi
^2}{16}}\right)$, there is:
\begin{align}
\liminf\limits_{N\to\infty}P_{Con}\left(G(N), k_0(N)\right)=0
\end{align}
where $P_{Con}\left(G(N), k_0(N)\right)$ denotes the probability for
the event that the network $G(N)$ with parameter $k_0(N)$ is
connected.
\end{theorem}
\begin{proof}
See Appendix \ref{appendix1}.
\end{proof}

Since $ \frac{\log\left(N+\pi/4 -\sqrt{\pi N/2+\pi
^2/16}\right)}{\log N}\to 1$ as $N\to\infty$, this theorem improves
the lower bound for the network connectivity from $0.074\log N$ to
$0.129\log N$.

\section{Conclusion}
In this paper, we improve the lower bound on the number of neighbors
required for the asymptotic connectivity of a dense ad hoc network.
Critical to the proof is the use of the $G^{Poisson}(N)$ model, for
which the distributions of nodes in non-overlapping areas are not
dependent.  The result is then extended from $G^{Poisson}(N)$ to the
$G(N)$ model of interest, resulting in an improvement in the lower
bound for the latter model to 0.129 $\log N$.


\appendices
\section{Proof of Theorem \ref{T: Poisson_not}}
\label{appendix1}
\begin{proof}
Since the convergence of $\left\{P_{Con}
\left(G\left(h(N)\right),k<0.129\log N \right), N=1,2,\dots\right\},
$ is unknown, ``$\liminf$'' is used instead of ``$\lim$'' throughout
the proof.  Naturally, we have:
\begin{align}
&\liminf\limits_{N\to \infty} P \left(G\left(N+\sqrt{\pi
N/2}\right)\text{ is connected  with }k<0.129\log N \right)\nonumber\\
 &\mbox{\hspace{0.15 cm}}= \liminf\limits_{M\to \infty} P
\left(G\left(M\right)\text{ is connected  with
}k<0.129\log \left(M+\frac{\pi}{4} -\sqrt{\frac{\pi M}{2}+\frac{\pi ^2}{16}}\right)\right) \label{approx6}\\
  &\mbox{\hspace{0.15 cm}}\leq \liminf\limits_{M\to \infty} P
\left(G\left(M\right)\text{ is connected  with
}k<0.129\log \left((M+1)+\frac{\pi}{4} -\sqrt{\frac{\pi (M+1)}{2}+\frac{\pi ^2}{16}}\right)\right)\label{approx7}\\
 &\mbox{\hspace{0.15 cm}}=  \liminf\limits_{N\to \infty} P
\left(G\left(\left(N\right)+\sqrt{\pi
\left(N\right)/2}-1\right)\text{ is connected  with }k<0.129\log N \right)\label{approx11}\\
 &\mbox{\hspace{0.15 cm}}\leq \liminf\limits_{M\to \infty} P \left(G\left(M\right)\text{ is
connected  with
}k<0.129\log \left(M+\frac{\pi}{4} +\sqrt{\frac{\pi M}{2}+\frac{\pi ^2}{16}}\right)\right)\label{approx8}\\
&=  \liminf\limits_{N\to \infty} P \left(G\left(N-\sqrt{\pi
N/2}\right)\text{ is connected with }k<0.129\log N \right)
\end{align}
Similarly, $ \forall h(N)\in\left[N-\sqrt{\pi N/2}, N+\sqrt{\pi
N/2}\right]$:
\begin{align}
\liminf\limits_{N\to \infty} &P \left(G\left(N+\sqrt{\pi
N/2}\right)\text{ is connected  with }k<0.129\log N \right)\nonumber\\
 &\leq \liminf\limits_{N\to \infty} P
\left(G\left(h(N)\right)\text{ is connected  with }k<0.129\log N
\right) \label{approx9}
\end{align}
Assume that:
\begin{align}
\liminf\limits_{N\to \infty} P \left(G\left(N+\sqrt{\pi
N/2}\right)\text{ is connected  with }k<0.129\log N \right) > 0
\label{asumption}
\end{align}
and seek a contradiction. Recalling (\ref{approx9}):
\begin{align}
\liminf\limits_{N\to \infty} &\frac{1}{2}P \left(G\left(N+\sqrt{\pi
N/2}\right)\text{ is connected  with }k<0.129\log N \right)\nonumber\\
 &< \liminf\limits_{N\to \infty} P
\left(G\left(h(N)\right)\text{ is connected  with }k<0.129\log N
\right)
\end{align}
Combined with (\ref{approx6}), (\ref{approx7}) and (\ref{approx11}),
this yields:
\begin{align}
&\liminf\limits_{M\to \infty} \frac{1}{2}P
\left(G\left(M\right)\text{ is connected  with }k<0.129\log
\left(M+\frac{\pi}{4} -\sqrt{\frac{\pi
M}{2}+\frac{\pi ^2}{16}}\right)\right)\nonumber\\
 &\mbox{\hspace{0.15 cm}}= \liminf\limits_{N\to
\infty} \frac{1}{2}P \left(G\left(N+\sqrt{\pi
N/2}\right)\text{ is connected  with }k<0.129\log N \right)\\
 &\mbox{\hspace{0.15 cm}}< \liminf\limits_{N\to \infty} P
\left(G\left(N+\sqrt{\pi N/2}-1\right)\text{ is connected with
}k<0.129\log N \right)\\
&\mbox{\hspace{0.15 cm}}= \liminf\limits_{M\to \infty} P
\left(G\left(M\right)\text{ is connected  with }k<0.129\log
\left((M+1)+\frac{\pi}{4} -\sqrt{\frac{\pi (M+1)}{2}+\frac{\pi
^2}{16}}\right)\right)
\end{align}
Then, it is obvious that: $\exists N_0$, such that $\forall N_1>N_0,
\forall N_2>N_0$,
\begin{align}
 &\inf\limits_{n_1\geq N_1} \frac{1}{2} P
\left(G\left(n_1\right)\text{ is connected  with
}k<0.129\log \left(n_1+\frac{\pi}{4} -\sqrt{\frac{\pi n_1}{2}+\frac{\pi ^2}{16}}\right)\right) \nonumber\\
 &\mbox{\hspace{0.35 cm}} < \inf\limits_{n_2\geq N_2} P
\left(G\left(n_2\right)\text{ is connected  with }k<0.129\log
\left((n_2+1)+\frac{\pi}{4} -\sqrt{\frac{\pi
(n_2+1)}{2}+\frac{\pi ^2}{16}}\right)\right)\\
 &\mbox{\hspace{0.35 cm}} \leq \inf\limits_{n_2\geq N_2}P
\left(G\left(n_2\right)\text{ is connected  with }k<0.129\log
\left(n_2+\frac{\pi}{4} +\sqrt{\frac{\pi n_2}{2}+\frac{\pi
^2}{16}}\right)\right)\label{approx10}
\end{align}
Thus, we have shown that there exists $N_0$, $\forall N>0$ such that
$N-\sqrt{\pi N/2}>N_0$,  for any $h(N)$ such that $
h(N)\in\left[N-\sqrt{\pi N/2}, N+\sqrt{\pi N/2}\right]$, there is:
\begin{align}
\inf\limits_{n\geq N} &\frac{1}{2}P \left ( G\left(n+\sqrt{\pi
n/2}\right)\text{ is connected  with }k<0.129\log n \right)\nonumber\\
 &< \inf\limits_{n\geq N}P \left(G\left(g(n)\right)\text{ is connected  with
}k<0.129\log n \right)
\end{align}
From Theorem \ref{T: Poisson}, let $k_0<0.129\log N$ and let $T$ be
the number of nodes in the square. Then,
\begin{align}
\lim\limits_{N\to \infty} P _{Con}(G^{Poisson}(N), k_0)
 &= \lim\limits_{N\to
\infty}\sum\limits_{j=0}^{\infty}P_{Con}\left(G(j), k_0\right)P(T=j)\\
 &= \lim\limits_{N\to
\infty}\sum\limits_{j=N-\sqrt{\pi N/2}}^{N+\sqrt{\pi
N/2}}P_{Con}(G(j),
k_0)P(T=j)\\
 &\geq \lim\limits_{N\to
\infty}\sum\limits_{j=N-\sqrt{\pi N/2}}^{N+\sqrt{\pi N/2}}
\inf\limits_{n\geq j}P_{Con}(G(n),
k_0)P(T=j)\\
 &> \lim\limits_{N\to
\infty}\sum\limits_{j=N-\sqrt{\pi N/2}}^{N+\sqrt{\pi
N/2}}\inf\limits_{n\geq N}
\frac{1}{2}P_{Con}\left(G\left(n+\sqrt{\pi n/2}\right),
k_0\right)P(T=j)\\
 &= \liminf\limits_{N\to \infty}\frac{1}{2}P_{Con}\left(G\left(N+\sqrt{\pi N/2}\right),
k_0\right)
\end{align}
Combined with (\ref{approx6}), this yields:
\begin{align}
0 &< \liminf\limits_{N\to \infty} P \left(G\left(N\right)\text{ is
connected  with }k<0.129\log \left(N+\frac{\pi}{4} -\sqrt{\frac{\pi
N}{2}+\frac{\pi ^2}{16}}\right)\right)
\nonumber\\
 &= \liminf\limits_{N\to \infty}P\left(G\left(N+\sqrt{\pi
N/2}\right) \mbox{ is connected with } k<0.129\log N\right)\\
 &< \liminf\limits_{N\to \infty} 2P (G^{Poisson}(N)
 \mbox{ is connected with } k<0.129\log N)
\end{align}
which contradicts  Theorem \ref{T: Poisson}. Thus, the assumption
(\ref{asumption}) is not correct.
\end{proof}


\begin{thebibliography}{1}

\bibitem{1Gupta and Kumar: Connectivity}
P. Gupta and P. R. Kumar, ``Critical power for asymptotic
connectivity in wireless networks,'' \emph{Stochastic Analysis,
Control, Optimization and Applications: A Volume in Honor of W.H.
Fleming}, 1998, edited by W.M. McEneany, G. Yin, and Q. Zhang,
(Eds.) Birkh¡§auser.

\bibitem{1Gupta and Kumar: Capacity}
P. Gupta and P. R. Kumar, ``The capacity of wireless networks,''
\emph{IEEE Trans. Inform. Theory}, vol. 46(2), pp. 388-404, March
2000.

\bibitem{1Xue and Kumar: Number}
F. Xue and P. R. Kumar, ``The number of neighbors needed for
connectivity of wireless networks,'' \emph{Wireless Networks}, vol
10, No. 2, pp.169-181, 2004.

\bibitem{1R. Meester and R. Roy: Continnum}
R. Meester and R. Roy, \emph{Continuum percolation}. Cambridge
University Press, 1996.

\bibitem{1Dousse: Connectivity of hybrid}
O. Dousse, P. Thiran, and M. Hasler, ``Connectivity in ad-hoc and
hybrid networks,'' in \emph{Proc. IEEE Infocom}, New York, June
2002.


\bibitem{1booth: percolation and geometry}
L. Booth, J. Bruck, M. Franceschetti, and R. Meester, ``Continuum
percolation and the geometry of wireless networks,'' \emph{Annals
of Applied Probability}, vol. 13(2), pp. 722-731, 2003.

\bibitem{1Dousse: con vs cap}
O. Dousse and P. Thiran, ``Connectivity vs Capacity in Dense Ad Hoc
Networks,'' in \emph{Proc. IEEE Infocom}, Hong Kong, Mar. 2004.

\bibitem{1Glouche: percolation threshold}
I. Glauche, W. Krause, R. Sollacher, M. Greiner, ``Continuum
percolation of wireless ad hoc communication networks''
\emph{Physica A} 325: 577 - 600, 2003.

\bibitem{1Hall: On continuum Percolation}
Peter Hall, ``On Continuum Percolation,'' \emph{The Annals of
Probability}, vol.13, No.4, pp. 1250-1266, Nov. 1985.


\bibitem{1Gilber: random networks}
E. N. Gilbert, ``Random plane networks,'' \emph{Journal of SIAM},
vol 9, pp. 533-543, 1961.


\bibitem{1Stauffer: intro to percolation}
D. Stauffer and A. Aharony, \emph{Introduction to Percolation
Theory} (Revised 2nd Edition). Taylor\&Francis, London, 1994.

\bibitem{1Penrose: geometry}
Mathew Penrose, \emph{Random Geometric Graph}. Oxford University
Press, 2003.

\bibitem{magic1}
L. Kleinrock and J.A. Silvester, ``Optimum transmission radii for
packet radio networks or why six is a magic number'', in \emph{Proc.
of IEEE Nat. Telecommun. Conf.}, pp. 4.3.1-4.3.5, Dec, 1978.

\bibitem{magic2}
H. Takagi and L. Kleinrock, ``Optimal transmission ranges for
randomly distributed packet radio terminals'', \emph{IEEE Trans.
Commun.}, vol 32, pp. 246-257, 1984.

\bibitem{magic3}
T. Hou and V. Li, ``Transmission range control in multihop packet
radio networks'', \emph{IEEE Trans. Commun.}, vol 34, pp. 38-44,
Jan. 1986.

\bibitem{magic4}
B. Hajek, ``Adaptive transmission strategies and routing in mobile
radio networks'', \emph{Proceedings of the Conference on Information
Sciences and Systems}, pp. 373-378,   Mar., 1983.

\bibitem{ours}
Sanquan Song, Dennis Goeckel and Don Towsley ,``Collaboration
Improves the Connectivity of Wireless Networks'' submitted to
\emph{Proc. IEEE Infocom}, Barcelona, April 2006.

\end{thebibliography}
\end{document}